# Combining Text Mining and Visualization Techniques to Study Teams' Behavioral Processes

Sherlock A. Licorish and Stephen G. MacDonell
*Department of Information Science*
*University of Otago*
PO Box 56, Dunedin 9054, New Zealand
sherlock.licorish@otago.ac.nz, stephen.macdonell@otago.ac.nz

**Abstract**

*There is growing interest in mining software repository data to understand, and predict, various aspects of team processes. In particular, text mining and natural-language processing (NLP) techniques have supported such efforts. Visualization may also supplement text mining to reveal unique multi-dimensional insights into software teams' behavioral processes. We demonstrate the utility of combining these approaches in this study. Future application of these methods to the study of teams' behavioral processes offers promise for both research and practice.*

**Keywords:** text mining; visualization; linguistic analysis

## 1. INTRODUCTION

A wealth of prior research has interrogated software repository data to examine and explain aspects of the software development process. In particular, given that the intricacies of team dynamics are evident in members' communication [1, 2], increasing effort has been directed to understanding teams' communication processes and relating these to within-project activities [3]. This is motivated by the understanding that good communication is essential for building positive interpersonal relations [4], an ingredient necessary for teams' productivity. Among the areas considered previously, research has revealed linkages between informal hierarchical communication structures and team performance [5], communication and coordination efficiency [6], and the quality of software [7]. Thus, studying the details in team communication and repository artifacts, and relating these to within-project phenomena, can provide valuable insights into the processes that are enacted during software development.

In this study we use psycholinguistics, text mining and visualization to examine repository data, so as to demonstrate the utility of combining these approaches to illuminate details of teams' behavioral processes evident in their artifacts. We first use self-organizing maps (SOMs) to depict various attributes of software tasks, both in isolation, and then as a collective, to assess their potential inter-relationships. We then leverage psycholinguistics to study practitioners' expression of behaviors, and correlate these to the aforementioned task attributes. We consider our findings in relation to previous work, we demonstrate the potential benefits of using text mining and visualization techniques to study repository artifacts, and we provide initial pointers for individuals undertaking software project governance.

In the next section we provide our study background and motivation, and we describe our research setting in Section 3. In Section 4 we present our results, and in Section 5 we summarize our study outcomes, and consider their potential implications.

## 2. BACKGROUND AND MOTIVATION

Prior research has investigated practitioners' artifacts to provide numerous insights into aspects of the software development process. Apart from the examination of software teams' communication patterns [8-10], works such as those of Bacchelli et al. [11] and Antoniol et al. [12] have assessed the utility of NLP techniques for extracting knowledge from email and bug description information. Bacchelli et al. [11] used regular expressions and other information retrieval approaches to examine how communication was related to source code changes and found that the analysis approach using regular expressions in emails outperformed more complex probabilistic and vector space models. Through the use of decision trees, naïve Bayes classifiers and logistic regression, Antoniol et al. [12] were also able to classify bugs based on specific terms used in the textual descriptions of such tasks. Similar outcomes are reported in other recent works [13, 14]. Visualization techniques have also been used to extract and depict knowledge about software systems. For example, Beyer and Hassan [15] provided Evolution Storyboard, a tool that uses animated panels to visualize the evolution of software structure. De Souza et al. [16] created the Ariadne plug-in for Eclipse to visualize dependencies among code components and developers. Wettel et al. [17] evaluated the adequacy of CodeCity, a 3D software visualization approach based on a city metaphor, for aiding program comprehension.

While text analysis methods have been used to understand aspects of software development [11, 13, 14], comparatively less effort has been dedicated to examining how software practitioners' behavioral processes relate to within-project activities [18-22]. Similarly, visualization techniques no doubt have utility beyond understanding properties of practitioners' code, and may be combined with text mining to enable the software engineering community to explain how software activities relate to teams' behavioral processes – knowledge that may not be easily detected using other methods. We address this gap, and so apply text mining and visualization techniques to demonstrate the utility of combining these approaches for revealing details in and about team artifacts.

## 3. RESEARCH SETTING

We examined development artifacts derived from release 1.0.1 of Jazz (based on the IBMR RationalR Team ConcertTM (RTC)1), a comprehensive software development and project management environment [23]. The Jazz environment supports work planning and traceability, software builds, code analysis, bug tracking, version control, and many other features [24]. Changes to source code in the Jazz environment are made via work items (WIs). The Jazz repository itself comprised product and process data collected from distributed teams undertaking software development and management across the USA, Canada and Europe. Teams have multiple roles (programmer, team lead, admin), and a project manager is responsible for the management and coordination of the activities undertaken by each team [25]. Jazz teams use the Eclipse-way methodology for guiding the development process [23]. This methodology outlines iteration cycles that last six to eight weeks, comprising planning, development and stabilizing phases, generally adhering to agile principles (albeit with longer iterations than are typical in agile contexts). All information for the software process is stored in the server repository, which is accessible through a web-based or Eclipse-based (RTC) client interface (refer to [3] for further details). Given Jazz's consolidated data capture and storage, we believe that insights gained by studying such artifacts would be more complete than those from other open source repositories [26].

We leveraged the IBM Rational Jazz Client API to extract team information along with development and communication artifacts from the repository. In total we extracted 30,646 resolved WIs (support tasks, defects and enhancements) developed across 30 iterations by a total of 474 practitioners working around these tasks between June 2005 and June 2008. Practitioners communicated 117,101 messages (comments) related to the 30,646 tasks (or WIs). We randomly selected a third of these tasks and associated comments for our analysis.

We would contend that the high volume of artifacts considered in this work would be adequate for drawing inferences; however, we do not make any claims for generalizability of our outcomes, as our primary focus in this study is to explore the benefits of using text mining and visualization techniques to study teams' behavioral processes from software repository artifacts. Accordingly, we use the task (WI) as the unit of analysis, and study multiple attributes of the task, relating these to practitioners' expressions of behaviors when working towards their resolution. We describe our measurements and techniques in the following two subsections.

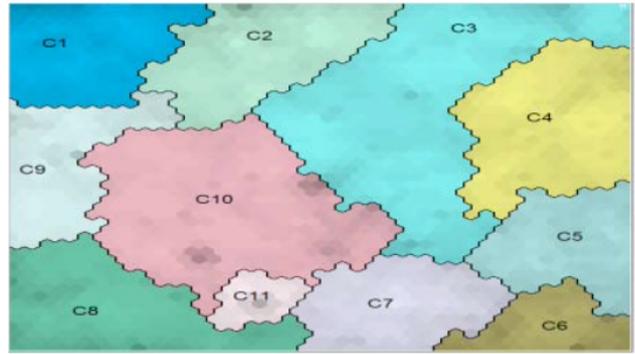

Fig. 1. Overall cluster map for task data.

### A. Text Mining and SOMs Visualization

In line with our goal of relating software practitioners' behavioral processes (evident in their textual communications) to task attributes, we sought an analysis approach capable of isolating measures while still considering the potential for relationships among them. We therefore first utilized the SOM unsupervised learning algorithm to reveal related patterns in pre-processed task data (e.g., task duration, priority, comment count). (Due to space constraints we refer the reader to [3] for details of our data pre-processing.) Kohonen's SOM employs unsupervised algorithmic training to classify multidimensional data into similarity graphs and clusters [27]. This procedure groups similar vectors (and clusters) based on their relative Euclidean distance using a nonparametric, recursive regression process [27]. We visualized the output of the SOM clustering through the Viscovery SOMine package (viscovery.net). This enabled us to project the clustered task (WI) data onto a 2D map of overlays, facilitating inspection and follow up statistical testing. Figure 1 provides an illustration of the output from the Viscovery SOMine package. It shows that our tasks were automatically grouped into eleven multi-dimensional clusters (C1 to C11) by the SOM algorithm (although the number of SOM clusters that is most analytically sensible may instead be specified by the user). We then repeated this modelling process for the linguistic measures used to assess practitioners' behaviors over the same tasks. This issue is considered next.

### B. Measuring Behaviors

Language use has been studied extensively across a range of social contexts [28-35]. These works have all provided evidence in support of the contention that there are unique variations in individuals' linguistic styles from situation to situation, and linguistic analysis of textual communication can reveal much about those who are communicating. In following the lead of previous work [22, 36], we employed the Linguistic Inquiry and Word Count (LIWC) software

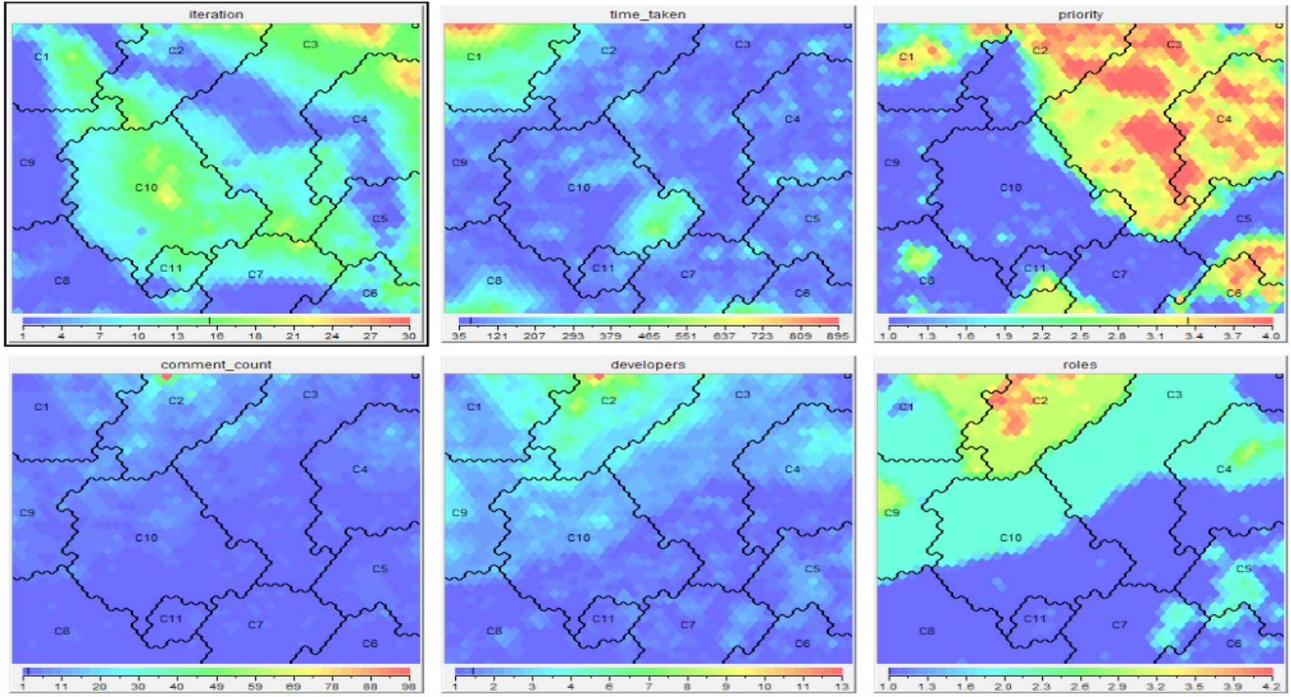

Fig. 2. Component maps for tasks' attributes.

TABLE I. KENDALL TAU-B CORRELATION (T) RESULTS

| Factor | 1 | 2 | 3 | 4 | 5 | 6 |
|---|---|---|---|---|---|---|
| 1 iteration | 1.0 | -0.20 | **0.30*** | 0.08 | 0.06 | 0.03 |
| 2 time taken | | 1.0 | -0.01 | 0.14 | 0.13 | 0.12 |
| 3 priority | | | 1.0 | 0.10 | 0.07 | 0.06 |
| 4 comment count | | | | 1.0 | **0.75*** | **0.58*** |
| 5 developer count | | | | | 1.0 | **0.75*** |
| 6 role count | | | | | | 1.0 |

Note: *$p < 0.05$; bold values represent noteworthy results

tool in our analysis of practitioners' behaviors. The LIWC tool was created after four decades of research using data collected across the USA, Canada and New Zealand [34]. This tool captures over 86% of the words used during conversations. Written text is submitted as input in a file that is then processed and summarized based on the LIWC tool's dictionary. Each word in the file is searched for in the dictionary, and specific type counts are incremented based on the associated word category (if found), after which a percentage value is calculated by aggregating the number of words in each linguistic category over all words in the messages. For example, if there were 10 instances of words belonging to the "social" dimension in a message with a length of 200 words then the percentage value for the "social" dimension would be (10/200=)5.0%. The dimensions in the LIWC output summary are said to capture the behaviors of individuals by assessing the words they use [34, 36]. We applied this tool to study various aspects of practitioners' and teams' behaviors previously [18, 19, 37, 38], although, we did not explore the utility of combining text mining and visualization techniques in these works. In the current study we consider six classes of behaviors that can be readily detected in language use: social, positive, negative, cognitive, work, and achievement behaviors. To illustrate, social behavior is indicated through the use of words such as "give", "buddy" and "love", while other words including "think", "consider" and "determine" reflect cognitive behavior [34]. We examine practitioners' expression of behaviors in relation to task attributes, using SOMs and statistics, as presented next.

## 4. RESULTS

Figure 2 depicts six SOMs representing various task attributes: iteration, time taken, priority, comment count, developers and roles. From the iteration map in Fig. 2 it is evident that most tasks (WIs) were created in the early and middle iterations (represented by blue and green colored areas), with far fewer tasks being created in later development cycles of the project (red areas of the map). In terms of the time taken to complete tasks, cluster C1 comprised tasks that took the longest, while most tasks were executed in around 35 days. In fact, in view of this evidence, and comparing this with the iteration map for C1, we can see that these tasks were typically created in the early and middle iterations. There were pockets of tasks in clusters C6, C8 and C10 that took longer than was typical, and these tasks were all created in the early and middle iterations. The priority map shows that tasks were typically raised with either low priority (1) or high priority (between 3 and 4). Additionally, in comparing the iteration and priority maps we can see that both low and high priority tasks were raised across iterations; however, tasks created in the later iterations tended to have slightly higher priorities. Interestingly, those tasks that took longer to complete in cluster C1 were mostly assigned medium to high priority. The comment count map shows that there were generally consistent contributions of messages around tasks for all iterations, with the exception of some tasks in cluster C2 (early and middle iterations) which generated more messages. In terms of developer involvement, Fig. 2 shows that more developers came together to solve a proportion of tasks that were scattered across the early and middle iterations. The spread of roles was similar; however, tasks developed in later iterations also involved many roles. Visualizing multiple task attributes through aligned SOMs

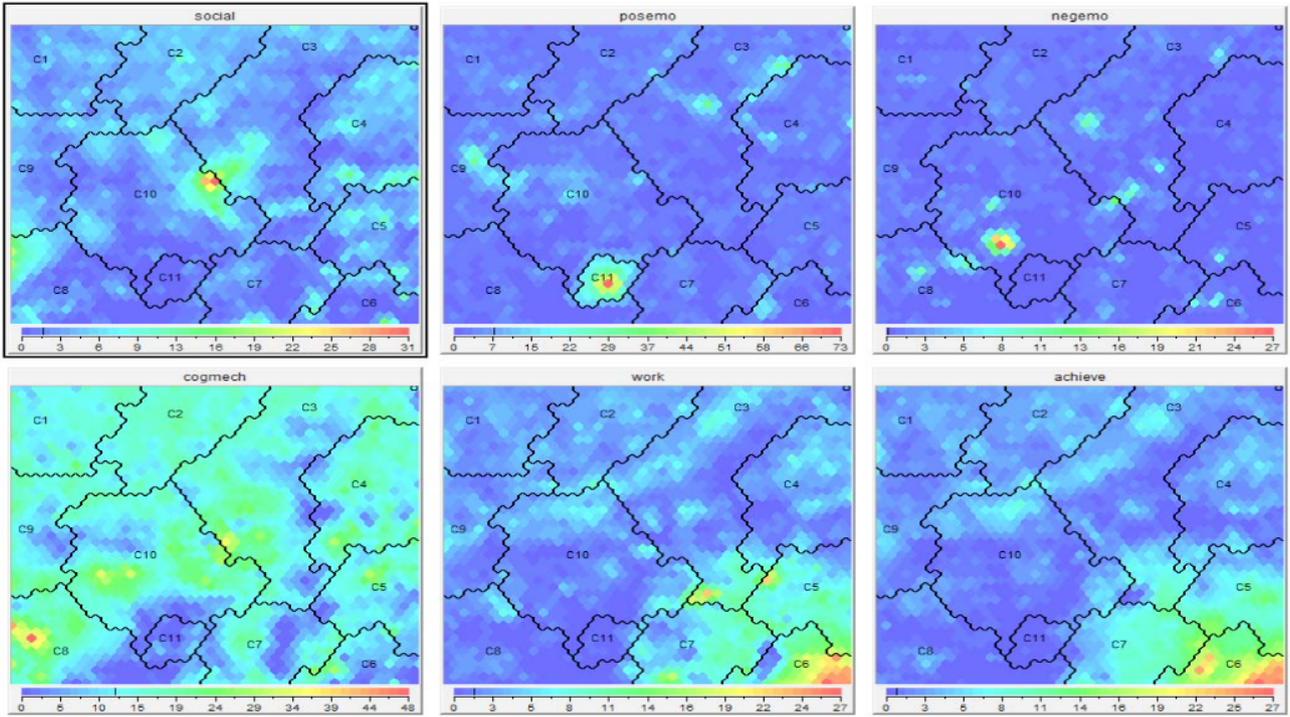

Fig. 3. Component maps for behaviors

is, we suggest, more informative than describing them in text.

We applied statistical testing to evaluate the significance of these results and to triangulate some of our informal visual SOM assessments. We first used the Kolmogorov-Smirnov test to check the normality of the distributions (e.g., for iteration, priority, and so on). The results of these tests confirmed that all the data significantly deviated from a normal distribution ($p < 0.05$). We then performed non-parametric Kendall tau-b correlation tests to examine to what extent the data were related (although only in terms of variable pairs), the results of which are provided in Table I. Of note in Table I is that software tasks were raised with higher priorities as the project progressed. In addition, but rather more intuitively obvious, the numbers of messages contributed increased with larger teams and a wider spread of roles, and the larger the cohort of practitioners working on software tasks the higher the number of roles that were involved.

We again utilized SOMs to cluster Jazz practitioners' expression of behaviors in relation to tasks. These results are depicted in the component maps of Fig. 3, which show that when resolving tasks members generally were consistently social. The posemo map in Fig. 3 reveals that members working on tasks clustered in C11 were extremely positive, while members working around some tasks in C10 tended to express more negative emotion (see negemo map). Cognitive language (as depicted in the cogmech map) was exchanged across most tasks over the 30 iterations, and particularly by team members working to resolve tasks clustered in C8 and C10. Finally, maps for work and achieve in Fig. 3 reveal that members working on a specific subset of tasks were extremely focused on work and achievement. These two maps show almost identical clustering. Kendall tau-b correlation tests were conducted to determine the extent of any bivariate relationships between the expression of behaviors and the six task attributes (i.e., iteration, priority, and so on). The most notable results are as follows: practitioners communicated significantly more social ($\tau = 0.30$, $p < 0.01$) and negative ($\tau = 0.33$, $p < 0.01$) utterances on tasks that had more messages, and there were significantly more negative utterances on tasks with larger teams, $\tau = 0.30$, $p < 0.01$. These were all moderate correlations. We summarize these findings and consider their implications for research and practice next.

## 5. SUMMARY AND IMPLICATIONS

Text mining and visualization can provide novel insights when applied to the study of teams' behavioral processes using repository data. Our visualizations revealed that most Jazz software tasks were created in the early and middle project iterations, with far fewer new tasks created in the latter project stages. Software tasks that were raised in the early and middle phases took longer than those that were coded towards project completion. Those governing software projects may leverage such patterns in terms of staffing strategies to sustain team productivity. In fact, a plausible explanation for the variance in time taken to resolve tasks across iterations could be task difficulty at project initiation, where the need to define a suitable software architecture is critical. Similarly, lower levels of schedule and release pressure in early iterations may mean practitioners work with less urgency. Such a finding may also be assessed in relation to team formation and maturity processes, as evidence has shown that teams are least functional at project start, where the need to test group structure and maintain expressive team concerns is higher [39]. This need decreases over time, when groups generally become more functional, tending to focus on task completion. Such assessments may explain the differences noted for task duration across iterations. That said, the evidence for the generally higher priorities assigned to tasks created in the latter iterations may have also mediated our results. We note too that it is understandable for release pressure to drive team urgency, and hence, the need for priorities to be consciously elevated in later project iterations. These are conceivable suppositions; insights from within-project experiences would enhance the perception of those governing software projects should they see similar evidence in artifacts visualizations.

Through our SOMs visualization and linguistic analysis we observed that, understandably, increases in the size of practitioner cohorts undertaking tasks were associated with higher communication volumes and increased role variation. Such evidence could have important implications regarding demands on communication channels and access to members, particularly in distributed development contexts. Of additional interest is our finding of heightened emotions (both positive and negative) among tasks involving higher numbers of developers. Perhaps positive emotion is used to offset negative emotion that is expressed due to some form of dissatisfaction? Overall we believe that the patterns noted in our analysis are interesting in terms of promoting thought-provoking research enquiries and informing software project governance.

## ACKNOWLEDGMENT

We thank IBM for granting us access to the Jazz repository. S. Licorish was supported by an AUT University Vice-Chancellor's Doctoral Scholarship Award.